\documentclass[twocolumn,showpacs,preprintnumbers,prb]{revtex4}
\usepackage{amssymb}
\usepackage[dvips]{color}
\usepackage{epsfig}
\usepackage{dcolumn}
\usepackage{bm}

\def\g{\gamma}
\def\G{\Gamma}
\def\d{\delta}

\def\e{\epsilon}

\def\t{\tau}

\def\s{\sigma}

\begin{document}

\title[ecp]{Spin-polarized current and shot noise in the presence of spin ﬂip in a quantum dot via nonequilibrium Green's functions}
\author{F. M. Souza,$^{1,2,4}$ A. P. Jauho,$^{2,3}$ and J. C. Egues$^4$}
\affiliation{$^1$International Centre for Condensed Matter
Physics, Universidade de Bras{\'i}lia, 70904-910, Bras{\'i}lia-DF,
Brazil \\ $^2$MIC - Department of Micro and
Nanotechnology, NanoDTU, Technical University of Denmark, {\O}rsteds
Plads, Bldg. 345E, DK-2800 Kgs. Lyngby, Denmark  \\ $^3$Laboratory
of Physics, Helsinki University of Technology, P. O. Box 1100,
FI-02015 HUT, Finland\\ $^4$Departamento de F\'{\i}sica e Inform\'{a}tica, Instituto de
F\'isica de S\~{a}o Carlos, Universidade de S\~ao Paulo, 13560-970
S\~ao Carlos, S\~ao Paulo, Brazil.}  \pacs{PACS number}

\begin{abstract}
Using non-equilibrium Green functions we calculate the
spin-polarized current and shot noise in a
ferromagnet--quantum-dot--ferromagnet (FM-QD-FM) system. Both
parallel (P) and antiparallel (AP) magnetic configurations are
considered. Coulomb interaction and coherent spin-flip (similar to a transverse magnetic field) are taken into
account within the dot. We find that the interplay between Coulomb
interaction and spin accumulation in the dot can result in a
bias-dependent current polarization $\wp$. In particular, $\wp$ can
be suppressed in the P alignment and enhanced in the AP case
depending on the bias voltage. The coherent spin-flip can also result in a switch of the
current polarization from the emitter to the collector lead.
Interestingly, for a particular set of parameters it is possible
to have a polarized current in the collector and an unpolarized
current in the emitter lead. We also found a suppression of the
Fano factor to values well below 0.5.
\end{abstract}

\volumeyear{year} \volumenumber{number} \issuenumber{number}
\eid{identifier}
\date[Date: ]{\today}
\maketitle

\section{INTRODUCTION}

Spin-dependent transport in quantum dots is a subject of intense
study nowdays due to its relevance to the new generation of proposed
spintronic devices that encompasses, for instance, the Datta-Das
transistor,\cite{dattadas} memory devices\cite{sc02,mk04} and as
an ultimate goal quantum
computers.\cite{dda02} In
particular, the recent progress in the coherent control of
electron spins in quantum dots
\cite{jme04,acj05,fhlk06} has
stimulated even further the research in this field, for possible
applications in quantum computation and quantum information
processing.\cite{man00} In addition to these fascinating technological
applications, quantum dots constitute a unique well-controllable
system to study fundamental physical aspects of transport in the
strong Coulomb-correlated regime, and its interplay with
spin-dependent effects.

A common geometry used for transport studies in quantum dots consists of two leads
weakly coupled to a quantum dot (QD) via tunneling barriers. Spin-dependent effects such as spin accumulation 
and spin-polarized transport can occur in these systems when both leads are (or at least one of them is) ferromagnetic (FM).
The junction FM-QD-FM resembles the standard TMR \cite{mj75,jsm95} and
GMR \cite{mnb88} geometries composed of an insulator layer
sandwiched by two ferromagnetic metallic leads, except for the
quantum dot replacing the insulator layer. This system (dot coupled to FM leads) was recently 
experimentally realized in the context of semiconductor quantum dots\cite{kh07_1,kh07_2} and molecules.\cite{anp04,ss05,cam07}
A wealth of novel spin-dependent effects has been observed in this system
due to the interplay of quantum confinement, Coulomb correlations,
Pauli principle and lead-polarization alignments. For instance, novel
effects such as spin-accumulation,\cite{iw06,iw05} spin-diode,\cite{fms07_1,commentNina} spin-blockade,\cite{fe06,ac04,ac04_2,mp03} 
spin-current ringing,\cite{fms07_3,ep08} negative differential
conductance and negative TMR,\cite{fe06,iw06} and so on, arise in
this context. In order to obtain additional information, not
contained in the average current, shot noise has also been analyzed in several
spintronic systems. A few exemples include shot noise in
spin-valve junctions\cite{brb99,yt01,egm03,mz05} and quantum dots attached to
ferromagnetic leads.\cite{rl03,fms02,du07}

Here we apply the Keldysh non-equilibrium technique to study
spin-polarized transport (current and shot noise) in a FM-QD-FM
system (Fig. 1). Both parallel (P) and antiparallel (AP) lead
magnetization alignments are considered. The left and the right
lead materials are taken to be different, thus resulting in additional
effects, not seen for leads with the same material. We analyze
both the current and the shot noise in the presence of Coulomb
interaction and spin-flip in the dot. We find an
interplay between spin accumulation and Coulomb interaction that
gives rise to a bias dependent current polarization $\wp$. More
specifically, $\wp$ can be suppressed or enhanced and have its
sign changed, depending on the magnetic alignment and the bias
voltage. We also note that the spin-flip can switch the current
polarization as it flows from the emitter to the collector lead.
In particular, it is possible to have an unpolarized
emitter-current and a polarized collector-current. For the shot noise, we find that spin-flip
can suppress it (in the AP case) with Fano factors reaching well
below 1/2.

The outline of our paper is as follows. In Sec. II we describe in
detail our model Hamiltonian. In Sec. III we present the
current and the noise calculations, respectively, including
general formulas for these quantities. In Sec. IV we present and
discuss numerical results for the current and the shot noise.
We summarize our conclusions in Sec. V. Technical details of our
calculation are described in the appendixes A-D.

\begin{figure}[tbp]
\par
\begin{center}
\epsfig{file=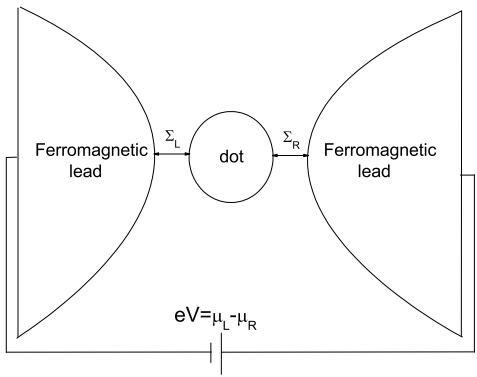, width=0.35\textwidth}
\end{center}
\caption{Schematic of the system studied: a quantum dot coupled to
two ferromagnetic leads. On the forward bias the electrons tunnel
from the left lead to the right lead via the
quantum dot. We consider configurations in which the FM leads are
parallel (P) or antiparallel (AP). The dot has a single orbital
level and can hold at most two electrons of opposite spins. Intradot Coulomb
interaction and spin-flip are considered.} \label{fig1}
\end{figure}

\section{Model system and Hamiltonian}

Our system consists of a quantum dot with one quantized level
coupled to two ferromagnetic leads via tunneling barriers. While the
electrons in the leads are noninteracting, the electrons in the dot
experience Coulomb repulsion and spin-flip scattering. The system
Hamiltonian is
\begin{equation}\label{H}
H=H_L+H_R+H_D+H_T.
\end{equation}
The first three terms in (\ref{H}) correspond to the three different
regions: left lead, right lead and dot. The last term $H_T$
hybridizes these three regions thus allowing electrons to tunnel
from one region to the other. This term gives rise to current in the
presence of a bias voltage.

More explicitly, we have for the ferromagnetic leads
\begin{equation}
H_\eta=\sum_{k\sigma}
\epsilon_{k\sigma\eta}c_{k\sigma\eta}^\dagger c_{k\sigma\eta},
\end{equation}
where $\epsilon_{k\sigma\eta}=\epsilon_{k \eta}+ (-1)^{\d_{\sigma \downarrow}} \Delta$
(Stoner Model) is the spin-dependent energy of the electron in
lead $\eta=(L,R)$, with the band spin splitting $\Delta$, $\sigma=\uparrow,\downarrow$ and $\d_{\uparrow \uparrow (\downarrow \downarrow)}=1$, $\d_{\downarrow \uparrow (\uparrow \downarrow)}=0$. The operator $c_{k \sigma\eta}$ $(c_{k \sigma\eta}^\dagger)$ destroys (creates) an electron with wave vector $k$ and spin $\sigma$ in lead $\eta$. The dot Hamiltonian is
\begin{equation}\label{hd}
H_D=\sum_\s \epsilon_\s d_\s^\dagger d_\s+ U n_\uparrow
n_\downarrow +R (d_\uparrow^\dagger d_\downarrow
+d_\downarrow^\dagger d_\uparrow), \label{HD}
\end{equation}
where $\epsilon_\s$ is the dot level and $d_\s$ $(d_\s^\dagger)$
annihilates (creates) an electron in the dot with spin $\sigma$.
Our model assumes a single spin-degenerate orbital level in the
dot, $\epsilon_{\uparrow}=\epsilon_{\downarrow}=\epsilon$. More
specifically, our dot can be singly occupied by an electron with
spin up or down or doubly occupied by two electrons with opposite
spins. We account for the Coulomb interaction in the dot via the
Hubbard term with correlation parameter $U$. We assume a linear
voltage drop across the system: $\epsilon=\epsilon_0-\frac{eV}{2}$, where $ e>0$, $V$ is the
applied voltage and $\epsilon_0$ is the dot level for $V=0$. The left $\mu_L$ and the right $\mu_R$ chemical
potentials are related by $\mu_L-\mu_R=eV$. Here we assume that $\mu_L$ 
is constant and defines the origin of the energy. For
positive bias ($ \mu_L>\mu_R$) the left lead is the electron
emitter and the right lead is the collector. The last term in (\ref{hd})
accounts for a coherent spin-flip in the dot.\cite{footR} This term can represent, e.g.,
a local transverse magnetic field that coherently rotates the electron spin, 
which can be experimentaly realized via ESR techniques\cite{he01}
or by the Hanle effect.\cite{hanle,mb05}

Instead of carrying out the calculations with the Hamiltonian
(\ref{HD}), we perform the following canonical
transformation,
\begin{equation}\label{canonical}
d_{\sigma}=\frac{1}{\sqrt{2}}\sum_{i=1,2}
(-1)^{i\delta_{\sigma\downarrow}}d_i.
\end{equation}
With (\ref{canonical}), the dot Hamiltonian becomes
\begin{equation}  \label{HD_transformed}
H_D=\sum_{i=1,2} [\epsilon_0+(-1)^{i}R] d_i^{\dagger} d_i +U n_1 n_2,
\end{equation}
where $n_i=d_i^\dagger d_i$. Note that in (\ref{HD_transformed}) the dot level
is split into two levels; $\epsilon_1=\epsilon_0-R$ and
$\epsilon_2=\epsilon_0+R$. We note that this canonical transformation rotates the 
spin quantization axis (e.g., to the direction of a local transverse magnetic field), 
thus replacing the spin-flip term by a diagonal term with a split level.

The tunneling Hamiltonian in (\ref{H}) is
\begin{equation}  \label{HT}
H_T=\sum_{k\sigma\eta} (t_{k\sigma}c_{k\sigma\eta}^\dagger d_\s +
t_{k\sigma}^* d_\s^\dagger c_{k\sigma\eta}),
\end{equation}
where the matrix element $t_{k\sigma}$ connects an electronic state in lead $%
\eta$ to one in the dot. Observe that the hopping process between
the leads and the dot is spin conserving, i.e., $t_{k\sigma}$ does
not mix different spin components. Applying the transformation
(\ref {canonical}) in (\ref{HT}) we find

\begin{equation}
H_T=\sum_{k\sigma\eta i}\frac{(-1)^{i\delta_{\sigma\downarrow}}}{\sqrt{2}}%
\{t_{k\sigma}^* d_i^\dagger
c_{k \sigma \eta}+t_{k\eta\sigma}c_{k \sigma \eta}^\dagger d_i\}.
\end{equation}

Next we calculate current and noise for the model described above.

\section{Current and Noise}
\label{avg-cur}

The current is calculated in the standard way from the
definition $I_L(t)=\langle \hat{I}_L(t) \rangle$ where
$\widehat{I} _L(t)=-e \dot{N}_L $ is the current operator, with
$N_L=\sum_{k\sigma}
c_{k\sigma L}^\dagger c_{k\sigma L}$ being the total number operator, and $%
\langle ... \rangle$ is a thermodynamic average. From the Heisenberg equation $\dot{N_L}=i[H,N_L]$, we find
\begin{equation}\label{curroper}
 \hat{I}_L(t)=i e \sum_{k \s} [ t_{k \s} c_{k \s \eta}^\dagger (t) d_\s(t) - t_{k \s}^* d_\s^\dagger(t) c_{k \s \eta}(t) ],
\end{equation}
which results in the following current expression\cite{ym92,apj94}
\begin{equation}\label{curr1}
I_L(t)=2e\mathrm{Re} \sum_{k\sigma} t_{k\sigma} i \langle
c_{kL\sigma}^\dagger (t) d_\s (t) \rangle.
\end{equation}

A similar expression holds for the right lead current $I_R=-e\langle \dot{N}%
_R \rangle$. Since we are in stationary regime we have simply $I_L=-I_R$. Using the canonical transformation in Eq. (\ref{curr1}) we obtain
\begin{equation}  \label{curr2}
I_L=2e\mathrm{Re} \sum_{k\sigma i} t_{k \sigma} \frac{%
(-1)^{i\delta_{\sigma\downarrow}}}{\sqrt{2}} G_{ik \sigma L}^<(t,t),
\end{equation}
where $G_{i k \sigma L}^<(t,t)=i\langle c_{kL\sigma}^\dagger (t) d_i
(t)\rangle$ is the lesser Green function, which is calculated via the Keldysh nonequilibrium technique.\cite{lvk65,hh96}

As a starting point we construct the complex time Green function
$G_{i k\sigma L}(\tau,\tau^{\prime})=-i\langle T_c d_i(\tau)
c_{k\sigma L}^\dagger (\tau^{\prime})\rangle$, where $T_c$ is the
contour time-ordering operator and $\tau$ and $\tau^{\prime}$ are
complex times running along a complex contour.\cite{lvk65,hh96}
Then we go from the Heisenberg to the interaction picture, by introducing the
$S$-matrix operator $S=e^{-i\int_c d\tau \widetilde{H}_T (\tau)}$.
Here the tilde means that $H_T$ is in the interaction
picture. After expanding $S$ we find,\cite{apj94}

\begin{eqnarray}  \label{GsksLtautau'}
&&G_{i k\sigma L} (\tau,\tau^{\prime})=t_{k\sigma}^* \sum_j \frac{%
(-1)^{j\delta_{\sigma\downarrow}}}{\sqrt{2}}
\nonumber\\&&\phantom{xxxxxxxx}\times\int_c d\tau_1
G_{ij}(\tau,\tau_1)g_{k\sigma L}(\tau_1,\tau^{\prime}),
\end{eqnarray}
where $G_{ij}(\tau,\tau_1)=-i\langle T_c d_i(\tau)
d_j^\dagger(\tau_1)\rangle $ and $g_{k\sigma
L}(\tau_1,\tau^{\prime})=-i\langle T_c \widetilde{c}_{k\sigma
L}(\tau_1) \widetilde{c}_{k\sigma L}^\dagger (\tau^{\prime})
\rangle$. Note that while $G_{ij}(\t,\t_1)$ is in the
Heisenberg picture, $g_{k \s L}(\t_1,\t')$ is in the interaction picture (denoted by the tilded operators). This
``separability'' of the interaction and Heisenberg pictures follows
from the assumption of noninteracting electrons in the leads. This allows us to put
the ``difficult'' part of the analysis entirely in the dot Green
functions, which contain the Coulomb interaction, the spin-flip
and the coupling to leads.

The next step is to apply Langreth's analytical continuation rules
\cite{hh96}
to (\ref{GsksLtautau'}), to find the lesser Green function appearing in (\ref{curr2}). This yields
\begin{eqnarray}
&&G_{ik \sigma L}^< (t,t^{\prime})=t_{kL\sigma}^* \sum_j \frac{%
(-1)^{j\delta_{\sigma\downarrow}}}{\sqrt{2}} \int dt_1\times  \nonumber \\
&&\{G_{ij}^r(t,t_1)g_{k \sigma L}^<(t_1,t^{\prime})+G_{ij}^<(t,t_1)
g_{k \sigma L}^a(t_1,t^{\prime})\},
\end{eqnarray}
where the labels $r$, $a$ and $<$ mean retarded, greater and
lesser, respectively. The calculation of the retarded $G_{ij}^r$, and lesser $%
G_{ij}^<$ dot Green functions is presented in the Appendix A.

Using this result in (\ref{curr2}) we arrive at
\begin{eqnarray}
I_L&=& 2e \mathrm{Re} \sum_{ij} \int dt_1  \nonumber \\
&\times&\{G_{ij}^r(t,t_1) \Sigma_{ji}^<(t_1,t) + G_{ij}^<(t,t_1)
\Sigma_{ji}^a(t_1,t) \}
\end{eqnarray}
where $\Sigma_{ji}^{L(<,a)}(t_1,t)$=$\frac{1}{2}\sum_{k\sigma}|t_{k%
\sigma}|^2 (-1)^{(i+j)\delta_{\sigma\downarrow}}$ $g_{k\sigma
L}^{(<,a)}(t_1,t)$, with the lesser Green function $g^<_{k\sigma
L}(t_1,t)=i\langle \widetilde{c}_{k\sigma L}^\dagger (t) \widetilde{c}%
_{k\sigma L}(t_1)\rangle$, and the advanced one $g_{k\sigma
L}^a(t_1,t)=i \theta(t-t_1) \langle \{\widetilde{c}_{k\sigma L}(t_1),
\widetilde{c}_{k\sigma L}^\dagger (t)\}\rangle$, here the curly
brackets denote an anticommutator.

In the steady state regime the Fourier transforms of the Green
functions result in single frequency Green functions. Since this
is the regime of interest here, we state for later use the Fourier
transform of the leads Green functions,
\begin{eqnarray}
g_{k\sigma \eta}^{a}(\e)&=&g_{k \sigma \eta}^{r^*}(\e)=\frac{1}{\e-\epsilon_{k\sigma \eta}- i\delta},\\
g_{k\sigma \eta}^<(\e)&=&2\pi i n_\eta(\e)\delta(\e-\epsilon_{k\sigma \eta}),
\end{eqnarray}
where $n_\eta$ is the Fermi distribution function of lead $\eta$.

\subsection{Average current in the stationary regime}

In a stationary regime all of the Green functions depend on only
$t-t_{1}$, yielding the Fourier transform
\begin{eqnarray}
I_{L}&=&2e\mathrm{Re}\int \frac{d\e }{2\pi }\sum_{ij}\{G_{ij}^{r}(\e
)\Sigma _{ji}^{L<}(\e )+G_{ij}^{<}(\e )\Sigma _{ji}^{La}(\e)\},\nonumber\\
&=&ie\int \frac{d\e }{2\pi }\mathrm{Tr}\{\mathbf{\Gamma }^{L}[(\mathbf{G}^{r}-\mathbf{G}^{a})n_{L}+\mathbf{G}^{<}]\},
\label{ILwrearranged}
\end{eqnarray}
with
\begin{equation}
\mathbf{\Gamma }^{L}=\frac{1}{2}\left(
\begin{array}{cc}
\Gamma _{\uparrow }^{L}+\Gamma _{\downarrow }^{L} & \Gamma
_{\uparrow
}^{L}-\Gamma _{\downarrow }^{L} \\
\Gamma _{\uparrow }^{L}-\Gamma _{\downarrow }^{L} & \Gamma
_{\uparrow
}^{L}+\Gamma _{\downarrow }^{L}%
\end{array}%
\right) ,  \label{GammaL}
\end{equation}%
where $\Gamma_{\sigma}^{L}=2\pi \sum_{k}|t_{k\sigma }|^{2}\delta
(\e -\epsilon _{k\sigma L})$ is the linewidth function. In what follows we neglect the energy-dependence
of $\G_\s^\eta$ (wideband limit), which will be taken as a constant phenomenological parameter.

\subsection{Spin-resolved currents}

From Eqs. (\ref{ILwrearranged}) and (\ref{GammaL}) we can also
determine the spin-resolved components of the average current
\begin{eqnarray}\label{ILs}
I_{L}^{\s} &=&ie\int \frac{d\e }{2\pi }\mathrm{Tr}\{\frac{%
\Gamma _{\s}^{L}}{2}\left(
\begin{array}{cc}
1 & (-1)^{\delta _{\sigma \downarrow }} \\
(-1)^{\delta _{\sigma \downarrow }} & 1%
\end{array}
\right)  \nonumber \\
&\times &[n_{L}(\mathbf{G}^{r}-\mathbf{G}^{a})+\mathbf{G}^{<}]\}.
\end{eqnarray}
A similar result holds for $I_R^\s$. Equation (\ref{ILs}) gives the spin-polarized current components
with their polarization axes defined along the magnetic moment of the leads. In the present study no spin-torque
is considered, which makes the projected current the relevant quantity to investigate. In the presence of spin-torque more general
definitions for spin-resolved charge currents and spin-currents should be used. A general expression for the spin-current in the presence of 
spin transfer was recently derived in Ref. [\onlinecite{mb05_2}].

\subsection{Noise definition}

Fluctuations of the current are interesting because they can give additional information about the
system beyond that provided by the average current
alone.\cite{ymb00} Here we derive an expression
for the current fluctuations, which include both thermal and shot
noise. The \textit{thermal noise} is related to fluctuations in
the occupations of the leads due to thermal excitation, and it vanishes at zero temperature.
\textit{Shot noise} is an unavoidable temporal fluctuation
of the current due to the granularity of the electron charge. It
is nonzero only for finite bias, i.e., it is a nonequilibrium
property. In the linear response regime the
fluctuation-dissipation theorem holds, yielding the relation
$S(\omega)=4k_B T G(\omega)$, where $G(\omega)$ is the
conductance.\cite{ymb00} Hence, in equilibrium the noise contains the same
information as the conductance. Away from equilibrium this
relation is no longer valid and the noise spectrum can provide
additional information.

We define noise via $S_{\eta\eta^{\prime}}=\langle \{\delta
\widehat{I}_\eta(t),\delta
\widehat{I}_{\eta^{\prime}}(t^{\prime})\}\rangle$, where $ \delta
\widehat{I}_\eta (t)=\widehat{I}_{\eta}(t)-I_{\eta}$ is the current
fluctuation at a time $t$ in lead $\eta$. Equivalently,
\begin{equation}  \label{Sdefinition}
S_{\eta\eta^{\prime}}(t,t^{\prime})=\langle \{\hat{I}_\eta(t),\hat{I}%
_{\eta^{\prime}}(t^{\prime})\}\rangle-2I_\eta^2,
\end{equation}
where we use the fact that $I_{\eta}=\langle \widehat{I}_\eta
(t)\rangle=\langle \widehat{I}_{\eta^{\prime}}(t^{\prime})\rangle$
in the stationary regime. Using the current operator
$\widehat{I}_\eta$ [Eq. (\ref{curroper})] and Eq. (\ref{canonical}) in Eq. (\ref{Sdefinition}), we obtain
\begin{eqnarray}  \label{Setaeta'operators}
S_{\eta\eta^{\prime}}(t,t^{\prime})&=&(ie)^2\sum_{kk^{\prime}\sigma\sigma^{%
\prime}ij} \frac{1}{2} (-1)^{i\delta_{\sigma\downarrow}}(-1)^{j\delta_{%
\sigma^{\prime}\downarrow}}  \nonumber \\
&&\{ t_{k\sigma} t_{k^{\prime}\sigma^{\prime}}\langle
c_{k\sigma\eta}^\dagger(t)
d_i(t)c_{k^{\prime}\sigma^{\prime}\eta^{\prime}}^\dagger(t^{\prime})
d_j(t^{\prime})\rangle  \nonumber \\
&& -t_{k\sigma} t_{k^{\prime}\sigma^{\prime}}^* \langle
c_{k\sigma\eta}^\dagger(t) d_i(t) d_j^\dagger(t^{\prime})
c_{k^{\prime}\sigma^{\prime}\eta^{\prime}}(t^{\prime})\rangle  \nonumber \\
&& - t_{k\sigma}^* t_{k^{\prime}\sigma^{\prime}} \langle
d_i^\dagger(t) c_{k\sigma\eta}(t)
c_{k^{\prime}\sigma^{\prime}\eta^{\prime}}^\dagger(t^{\prime})d_j(t^{%
\prime})\rangle  \nonumber \\
&& + t_{k\sigma}^* t_{k^{\prime}\sigma^{\prime}}^*\langle
d_i^\dagger(t) c_{k\sigma\eta}(t) d_j^\dagger(t^{\prime})
c_{k^{\prime}\sigma^{\prime}\eta^{\prime}}(t^{\prime})\rangle\}  \nonumber \\
&&+h.c.-2I_\eta^2.
\end{eqnarray}

\subsection{Noise in terms of Green functions}

Each $\langle ... \rangle$ term in (\ref{Setaeta'operators}) can
be expressed in terms of a Green function. Defining the two particle Green functions,
\begin{eqnarray}
g^{(2)}_1(\tau,\tau^{\prime})&=&i^2 \langle T_c
c_{k\sigma\eta}^{\dagger}(\tau)d_i(\tau)c_{k^{\prime}\sigma^{\prime}\eta^{
\prime}}^{\dagger}(\tau^{\prime})d_j(\tau^{\prime})\rangle  \nonumber \\
g^{(2)}_2(\tau,\tau^{\prime})&=&i^2 \langle T_c
c_{k\sigma\eta}^{\dagger}(\tau)d_i(\tau)d_j^\dagger(\tau^{\prime})c_{k^{
\prime}\sigma^{\prime}\eta^{\prime}}(\tau^{\prime})\rangle  \nonumber \\
g^{(2)}_3(\tau,\tau^{\prime})&=&i^2 \langle T_c
d_i^\dagger(\tau)c_{k\sigma\eta}(\tau)
c_{k^{\prime}\sigma^{\prime}\eta^{\prime}}^{\dagger}(\tau^{\prime})d_j(
\tau^{\prime}) \rangle  \nonumber \\
g^{(2)}_4(\tau,\tau^{\prime})&=&i^2\langle T_c
d_i^\dagger(\tau)c_{k\sigma\eta}(\tau)d_j^\dagger(\tau^{\prime})c_{k^{
\prime}\sigma^{\prime}\eta^{\prime}}(\tau^{\prime})\rangle,
\nonumber
\end{eqnarray}
we can write (\ref{Setaeta'operators}) as
\begin{eqnarray}  \label{Setaeta'middle}
S_{\eta\eta^{\prime}}(t,t^{\prime})&=&
e^2\sum_{kk^{\prime}\sigma\sigma^{\prime}i j} \frac{1}{2}
(-1)^{i\delta_{
\sigma\downarrow}}(-1)^{j\delta_{\sigma^{\prime}\downarrow}}  \nonumber \\
&&\{ t_{k\sigma}
t_{k^{\prime}\sigma^{\prime}}g^{(2)>}_1(t,t^{\prime})
\nonumber \\
&& -t_{k\sigma} t_{k^{\prime}\sigma^{\prime}}^*
g^{(2)>}_2(t,t^{\prime})
\nonumber \\
&& - t_{k\sigma}^* t_{k^{\prime}\sigma^{\prime}}
g^{(2)>}_3(t,t^{\prime})
\nonumber \\
&& + t_{k\sigma}^* t_{k^{\prime}\sigma^{\prime}}^*
g^{(2)>}_4(t,t^{\prime})\}
\nonumber \\
&&+h.c.-2I_\eta^2,
\end{eqnarray}
where $g_{i}^{(2)>} (t,t^{\prime})$ is obtained from the complex time
Green function $g^{(2)}_i (\tau,\tau^{\prime})$ via analytical
continuation. Similarly to the current calculation where we
develop an \textit{S}-matrix
expansions in $G_{\sigma k\sigma L}(\tau,\tau^{\prime})$ to obtain $%
G^<_{\sigma k\sigma L}(t,t^{\prime})$, here we expand the $S$-matrix in $%
g^{(2)}_i (\tau,\tau^{\prime})$ and then obtain $g^{(2)>}_i
(t,t^{\prime})$. This procedure follows the standard calculations proposed in Ref. [\onlinecite{apj94}] to derive the
current equation. The details of this \textit{S}-matrix expansion
are presented in Appendix B; here we simply state the results,
\begin{eqnarray}\label{g1}
&&g^{(2)}_1(\tau,\tau^{\prime})=\frac{1}{2}\sum_{i_1,i_2=1,2}(-1)^{i_1
\delta_{\sigma\downarrow}} (-1)^{i_2
\delta_{\sigma^{\prime}\downarrow}}
\nonumber \\
&& \times t_{k\sigma}^*t_{k^{\prime}\sigma^{\prime}}^* \int \int
d\tau_1 d\tau_2 g_{k\sigma\eta}(\tau_1,\tau)
g_{k^{\prime}\sigma^{\prime}\eta^{\prime}}(\tau_2,\tau^{\prime})
\nonumber
\\
&& \{G_{i i_1}(\tau,\tau_1)G_{j i_2}(\tau^{\prime},\tau_2)-G_{i
i_2}(\tau,\tau_2)G_{j i_1}(\tau^{\prime},\tau_1)\},\nonumber \\
&&
\end{eqnarray}
and
\begin{eqnarray}  \label{g2}
&&g^{(2)}_2(\tau,\tau^{\prime})=-\delta_{k\sigma\eta,k^{\prime}\sigma^{%
\prime}\eta^{\prime}}g_{k\sigma\eta}(\tau^{\prime},\tau)G_{ij}(\tau,\tau^{%
\prime})  \nonumber \\ && \phantom{xxx} + \frac{1}{2}\sum_{i_1,i_2=1,2}(-1)^{i_1
\delta_{\sigma\downarrow}} (-1)^{i_2
\delta_{\sigma^{\prime}\downarrow}}
\nonumber \\  
&&  \times t_{k\sigma}^* t_{k^{\prime}\sigma^{\prime}}\int \int d\tau_1
d\tau_2 g_{k\sigma\eta}(\tau_1,\tau)
g_{k^{\prime}\sigma^{\prime}\eta^{\prime}}(\tau^{\prime},\tau_2)
\nonumber
\\
&&\{G_{i i_1}(\tau,\tau_1)G_{i_1 j}(\tau_2,
\tau^{\prime})-G_{i j}(\tau,\tau^{\prime})G_{i_2 i_1}(\tau_2,\tau_1)\}.  \nonumber \\
&&
\end{eqnarray}
Equations (\ref{g1}) and (\ref{g2}) hold on a Hartree-Fock or other mean-field theory (see details in Appendix B). The other two Green functions $g_3^{(2)}$ and
$g_4^{(2)}$ are given by
\begin{equation}  \label{g3}
g^{(2)}_3(\tau,\tau^{\prime})=g^{(2)^*}_2(\tau,\tau^{\prime}),
\end{equation}
and
\begin{equation}  \label{g4}
g^{(2)}_4(\tau,\tau^{\prime})=g^{(2)^*}_1(\tau,\tau^{\prime}).
\end{equation}
From a diagrammatic point of view the terms in Eqs. (\ref{g1}) and (\ref%
{g2}) involving
\begin{equation}
G_{i i_1}(\tau,\tau_1) g_{k\sigma\eta}(\tau_1,\tau) G_{j
i_2}(\tau^{\prime},\tau_2)g_{k^{\prime}\sigma^{\prime}\eta^{\prime}}(\tau_2,\tau^{\prime}),
\end{equation}
and
\begin{equation}
G_{\sigma\sigma}(\tau,\tau_1)g_{k\sigma\eta}(\tau_1,\tau) g_{k^{\prime}\sigma^{\prime}\eta^{\prime}}(\tau^{\prime},\tau_2)G_{\sigma^{\prime}\sigma^{\prime}}(\tau_2,%
\tau^{\prime})
\end{equation}
are disconnected. These disconnected terms, together with similar
ones in the equations for $g^{(2)}_3(\tau,\tau^{\prime})$ and $g^{(2)}_4(\tau,\tau^{\prime})$, cancel identically the term
$2I_{\eta}^2$ in Eq. (\ref{Setaeta'middle}) (see Appendix C). So we can say that this corresponds to the linked cluster expansion to the noise. 
The other terms in Eqs. (\ref{g1})-(\ref{g2}) give the connected diagrams and thus can give a
contribution to the noise. Substituting the connected terms of
Eqs.(\ref{g1})--(\ref{g4}) in Eq.(\ref{Setaeta'middle}) we find
\begin{eqnarray}  \label{Sttau}
&&S_{\eta\eta^{\prime}}(t,t^{\prime})=e^2\sum_{k\sigma}|t_{k\sigma}|^2\{%
\delta_{\eta,\eta^{\prime}} \sum_{ij}
(-1)^{(i+j)\delta_{\sigma\downarrow}}
\nonumber \\
&&\times[g_{k\sigma\eta}^>(t,t^{\prime})G_{ji}^<(t^{%
\prime},t)+G_{ij}^>(t,t^{\prime})g_{k\sigma\eta}^<(t^{\prime},t)]
\nonumber
\\
&&
-e^2\sum_{kk^{\prime}\sigma\sigma^{\prime}}|t_{k\sigma}|^2|t_{k^{\prime}%
\sigma^{\prime}}|^2\int_c \int_c d\tau_1 d\tau_2  \nonumber \\
&& \times \frac{1}{4} \sum_{i,j,i_1,i_2=1,2}
(-1)^{(i+i_1)\delta_{\sigma\downarrow}}
(-1)^{(j+i_2)\delta_{\sigma^{\prime}\downarrow}}  \nonumber \\
&& \times\{
G_{ii_2}(t,\tau_2)g_{k^{\prime}\sigma^{\prime}\eta^{\prime}}(\tau_2,t^{%
\prime})G_{ji_1}(t^{\prime},\tau_1)g_{k\sigma\eta}(\tau_1,t)  \nonumber \\
&&
-G_{ij}(t,t^{\prime})g_{k^{\prime}\sigma^{\prime}\eta^{\prime}}(t^{\prime},%
\tau_2)G_{i_2i_1}(\tau_2,\tau_1)g_{k\sigma\eta}(\tau_1,t)  \nonumber \\
&&
-g_{k\sigma\eta}(t,\tau_1)G_{i_1i_2}(\tau_1,\tau_2)g_{k^{\prime}\sigma^{%
\prime}\eta^{\prime}}(\tau_2,t^{\prime})G_{ji}(t^{\prime},t)  \nonumber \\
&&
+g_{k\sigma\eta}(t,\tau_1)G_{i_1j}(\tau_1,t^{\prime})g_{k^{\prime}\sigma^{%
\prime}\eta^{\prime}}(t^{\prime},t_2)G_{i_2i}(\tau_2,t)\}^{t>t^{\prime}}
\nonumber \\
&& \phantom{xxxxxxxxxxxxxxxxxxxxxxxxxx}+h.c.,
\end{eqnarray}
where the superscript $t>t^{\prime}$ means that
an analytical continuation should be performed by applying Langreth's rules.

\subsection{Zero-frequency shot noise}

The shot noise is defined as the Fourier transform of $S_{\eta\eta^{\prime}}(t-t^{\prime})$, which
in the stationary regime reads
\begin{equation}\label{noiseDefin}
S_{\eta\eta^{\prime}}(\omega)=\int_{-\infty}^{\infty}
d(t-t^{\prime})
e^{i\omega(t-t^{\prime})}S_{\eta\eta^{\prime}}(t-t^{\prime}).
\end{equation}
Using the analytical continuation of Eq. (\ref{Sttau}) into Eq.
(\ref{noiseDefin}) we find the following zero-frequency shot
noise,\cite{jxz03}
\begin{eqnarray} \label{SLL}
&&S_{LL}(\omega=0)=\frac{e^2}{\pi}\int d\e \mathrm{Tr} \{i n_L\mathbf{%
\Gamma}^L\mathbf{G}^>  \nonumber \\
&-&i(1-n_L)\mathbf{\Gamma}^L \mathbf{G}^<+\mathbf{\Gamma}^L \mathbf{G}^>\mathbf{\Gamma}^L%
\mathbf{G}^< \nonumber \\
&+& \mathbf{\Gamma}^L (\mathbf{G}^r-\mathbf{G}^a)\mathbf{\Gamma}^L (n_L\mathbf{G}^>-(1-n_L)%
\mathbf{G}^<) \nonumber \\
&-&n_L(1-n_L)(\mathbf{G}^a\mathbf{\Gamma}^L\mathbf{G}^a\mathbf{\Gamma}^L+%
\mathbf{G}^r\mathbf{\Gamma}^L\mathbf{G}^r\mathbf{\Gamma}^L) \},
\end{eqnarray}
where $\mathbf{G}^>$ satisfies the identity $\mathbf{G}^>-\mathbf{G}^<=%
\mathbf{G}^r-\mathbf{G}^a$. All the Green functions in Eq.
(\ref{SLL}) are in the frequency domain. In our analysis we take
only the component $\eta=\eta'=L$. Since the $dc$ noise is
position independent, we have simply
$S_{LL}(0)=S_{RR}(0)=-S_{LR}(0)=-S_{RL}(0)$. Equation (\ref{SLL}) can be expressed in a standard form
as follow\cite{lesovik,mb90} (Appendix D),
\begin{eqnarray} \label{SLL_buttiker}
&&S_{LL}(\omega=0)=\frac{e^2}{\pi}\int d\e
\nonumber\\&&\phantom{xxxx} \times \mathrm{Tr} \{
[n_L(1-n_L)+n_R(1-n_R)] \mathbf{T}(\e)
\nonumber\\&&\phantom{xxxxxxxxx}+ (n_L-n_R)^2 \mathbf{T}(\e)
[1-\mathbf{T}(\e)] \},
\end{eqnarray}
with the transmission matrix $\mathbf{T}=\mathbf{\Gamma}^L\mathbf{G}^r\mathbf{\Gamma}^R
\mathbf{G}^a$. In the calculation leading to Eqs. (\ref{SLL}) or (\ref{SLL_buttiker}) we have truncated an S-matrix
expansion by breaking two-particle Green function into products of one-particle Green functions. This procedure holds
in a mean field theory. Thus for a consistent application of Eqs. (\ref{SLL})-(\ref{SLL_buttiker}) a similar
approximation (Hartree-Fock like) for the Green functions should be made (see Appendix A). 
Some limitations imposed by this approximation are discussed in the end of Sec. \ref{results}.\cite{diagram}

\subsection{Model for the FM leads}

The ferromagnetism of the leads is considered via the spin-dependent parameter $\G_\s^\eta$.
From the Stoner model, for instance, we can see that the density of states for spin up electrons of the lead is shifted with respect to that of
the spin down electrons. Since $\G_\s^\eta$ contains information about the 
spin-dependent density of states it is expected that $\Gamma_\uparrow^\eta \neq \Gamma_\downarrow^\eta$.\cite{fms04} Following
Ref. [\onlinecite{wr01}] we define $\Gamma_\s^L=\Gamma_0[1+(-1)^{\delta_{\sigma \downarrow}} p_L]$. The parameter $\Gamma_0$ gives
the strength of the lead-to-dot coupling and $p_L$ is a parameter
describing the degree of spin polarization of the left lead.\cite{expvalues} Note that $%
\Gamma_\uparrow^L > \Gamma_\downarrow^L$ for $p_L \neq 0$. This
means that the population for spin up around the Fermi energy in the
left lead is greater than the population for spin down. Similarly,
for the right ferromagnetic lead we assume $\Gamma_\s^R=\Gamma_0 [1 +
(-1)^{\delta_{\sigma \downarrow}} p_R]$ for the parallel (P) and $\Gamma_\s^R=\Gamma_0 [1 - (-1)^{\delta_{\sigma \downarrow}} p_R]$ for the antiparallel (AP) lead alignments. Note that for the P case we have $\Gamma_\s^L=\Gamma_\s^R$ and for the AP
configuration $\Gamma_\s^L=\Gamma_{\overline{\sigma}}^R$, with
$\overline{\sigma}$ being the opposite of $\sigma$. In the present work we mostly discuss the $p_L \neq p_R$ case, i.e., 
a geometry in which the left and right leads are composed of different
materials (Ni and Co, for instance).

\subsection{Numerical procedure}

The numerical results are obtained following a self-consistent procedure. We calculate the average
\begin{equation}
 \langle \hat{n}_{ij} \rangle = \langle d_i^\dagger d_j \rangle = \int \frac{d\e}{2\pi i} G_{j i}^<(\e),
\end{equation}
self-consistently with Eqs. (\ref{A9})-(\ref{A15}). When converged solutions for the expectation values $\langle d_i^\dagger d_j \rangle$ 
and dot Green functions are found, we determine the current [Eq. (\ref{ILs})] and the noise [Eq. (\ref{SLL})]. This iterative schema is performed for each bias voltage. 

\section{RESULTS}\label{results}

\subsection{Spin resolved electronic occupations}

\begin{figure}[tbp]
\begin{center}
\epsfig{file=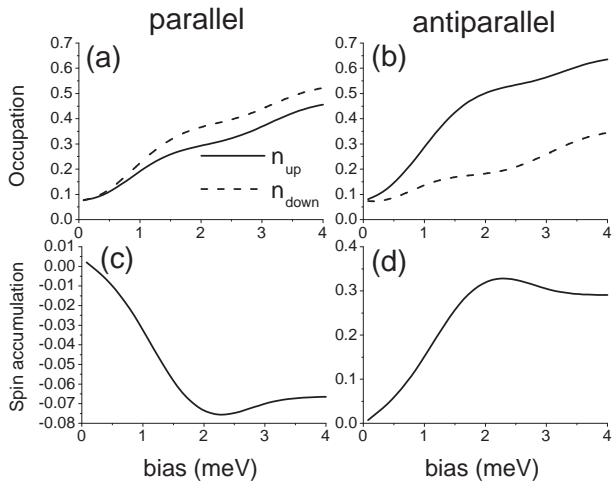, width=0.45\textwidth}
\end{center}
\caption{Spin-resolved occupations $n_\uparrow$ and $n_\downarrow$
and the spin accumulation, $m=n_\uparrow - n_\downarrow$, against
bias voltage in both parallel and antiparallel FM lead alignments.
In the P case $n_\uparrow < n_\downarrow$ while in the AP
alignment $n_\uparrow > n_\downarrow$. These imbalances give rise
to spin-dependent population suppressions in the bias range 1-3
meV. This translates into a tunable $m$ with the bias voltage. Parameters:
$p_L=0.23$, $p_R=0.35$, $k_B T=200$ $\mu$eV, $\G_0=10$ $\mu$eV,
$\e_0=0.5$ meV and $U=1$ meV.}\label{fig2}
\end{figure}

\emph{Occupations}. Figures \ref{fig2}(a)-(b) show the spin up and
spin down occupations of the dot for both parallel (P) and
antiparallel (AP) alignments. In the P case the dot has a net
spin-down polarization with $n_{\uparrow }<n_{\downarrow }$, while
in the AP configuration $n_{\uparrow }>n_{\downarrow }$. These
spin imbalances in the dot can be easily understood in terms of
the tunneling rates $\Gamma _{\s}^{\eta }$ adopted. The parameters
$p_L=0.23$ and $p_R=0.35$ used here\cite{parameters} yield the
following tunneling rates in the parallel case: $\Gamma _{\uparrow
}^{L}=12.3$ $\mu $eV, $\Gamma _{\downarrow }^{L}=7.7$ $\mu $eV,
$\Gamma
_{\uparrow }^{R}=13.5$ $\mu $eV and $\Gamma _{\downarrow }^{R}=6.5$ $\mu $%
eV. In the AP case the values of the tunneling rates to the right
lead are swapped ($\Gamma _{\uparrow }^{R}\rightleftarrows \Gamma
_{\downarrow }^{R}$). From these rates we conclude that in the P
case a spin up electron leaves the dot faster than it comes in.
The opposite happens for a spin down electron. The imbalance of
these in/out tunneling rates results in a larger spin down
occupation in the parallel case, i.e., $n_{\uparrow
}<n_{\downarrow }$, Fig. \ref{fig2}(a). By the same token, in the
AP alignment we have $n_{\uparrow }>n_{\downarrow }$ as seen in
Fig. \ref{fig2}(b).

\emph{Spin accumulation}. In Fig. \ref{fig2}(c)-(d) we show the
spin accumulation ($m=n_\uparrow - n_\downarrow$) as a function of
the bias voltage. In the zero bias limit $m$ is essentially zero.
When the bias increases the spin accumulation in the P case assume
negative values. In contrast, in the AP alignment $m$ is enhanced. In
particular, in the bias range corresponding to a singly-occupied dot
(1-3 meV)\cite{singleocup} the additional suppression [Fig.
\ref{fig2}(c)] or the enhancement [Fig. \ref{fig2}(d)] of $m$ is
due to the {\it spin-dependent population suppression} that takes
place in the presence of Coulomb interaction and spin
accumulation. More specifically, in the AP case due to Coulomb
interaction $n_{\uparrow }$ tends to suppress more strongly
$n_{\downarrow }$ than otherwise. This translates into an enhancement
of $m$. In the P alignment the spin up occupation $n_\uparrow$ is
more suppressed than $n_\downarrow$, thus $m$ becomes more
negative [Fig. \ref{fig2}(c)]. We emphasize that this effect
happens for both the P and AP alignments because we assume $p_L
\neq p_R$. For equal leads we find $n_\uparrow = n_\downarrow$ in
the P case so that $m$ remains zero in this configuration.

\subsection{Current and its polarization}

\begin{figure}[tbp]
\par
\begin{center}
\epsfig{file=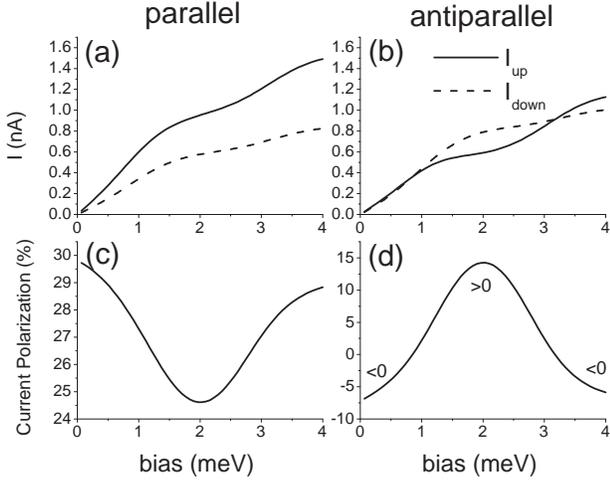, width=0.45\textwidth}
\end{center}
\caption{(a)-(b) Spin-resolved currents $I_\uparrow$ and
$I_\downarrow$ and (c)-(d) current polarization for both parallel
and antiparallel alignments. In the P case we find $I_\uparrow
> I_\downarrow$. In contrast, for AP alignment $I_\uparrow < I_\downarrow$ in the bias window $1-3$
meV (singly occupied regime) and $I_\uparrow > I_\downarrow$ when
the bias exceeds the charging energy above $3$ meV (doubly occupied
regime). Panels (c)-(d) reveal a suppression and an enhancement of
the current polarization in the singly occupied regime for the P
and AP cases, respectively. Parameters: $p_L=0.23$, $p_R=0.35$,
$k_B T=200$ $\mu$eV, $\G_0=10$ $\mu$eV, $R=0$, $\e_0=0.5$ meV and
$U=1$ meV.}\label{fig3}
\end{figure}

\emph{Current}. Figures \ref{fig3}(a)-(b) show the current in the
P and AP cases for $U=1$ meV and $R=0$. Similarly to the
occupations, some features of the spin-up and spin down currents
can be understood in terms of the tunneling rates. For instance,
their saturation values (second plateau) can be easily calculated
from the standard expresssion\cite{standardeq}
\begin{equation}
I^{\sigma}_\eta=e\frac{\Gamma _{\sigma}^{L}\Gamma _{\sigma}^{R}}{%
\Gamma _{\sigma}^{L}+\Gamma _{\sigma}^{R}},  \label{Issimple}
\end{equation}%
which gives $I^{\uparrow }_\eta>I^{\downarrow}_\eta$ and $I^{\uparrow
}_\eta<I^{\downarrow}_\eta$ in the P and AP cases, respectively. For the
first plateau Eq. (\ref{Issimple}) is not valid and these
inequalities can change.\cite{at03}

In the P case [Fig. \ref{fig3}(c)] $I_\uparrow$ is more strongly suppressed than
$I_\downarrow$ due to the interplay of spin accumulation ($n_{\uparrow} < n_{\downarrow
}$) and Coulomb interaction. This results in a supression of the current polarization [$\wp =(I_{\uparrow
}-I_{\downarrow })/(I_{\uparrow }+I_{\downarrow})$] in the range 1-3 meV. On the other hand, 
in the AP case [Fig. \ref{fig3}(d)] $I^\downarrow_\eta$ is more suppressed than $I^\uparrow_\eta$ due to
the inverted inequality $n_{\uparrow} > n_{\downarrow}$, thus resulting in an enhancement of $\wp$.

\begin{figure}[tbp]
\par
\begin{center}
\epsfig{file=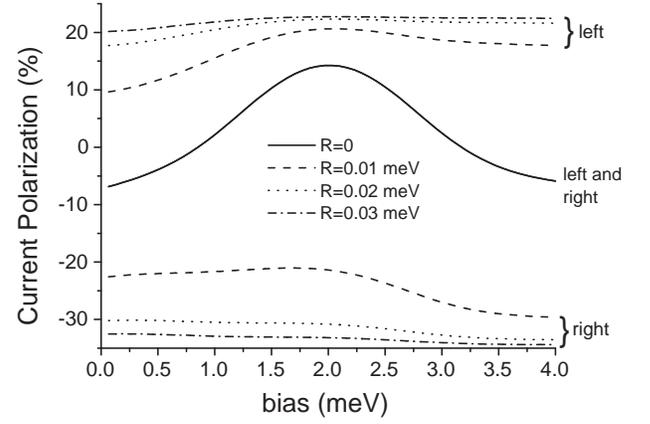, width=0.45\textwidth}
\end{center}
\caption{Current polarization against the bias voltage for differing
spin-flip rates in the AP alignment and in both left and right leads. For
$R=0$ the polarization is the same on both sides. When $R \neq 0$, though, the current polarization in the left side enhances,
tending to the left magnetization degree $p_L$.
In contrast, the current polarization in the right side assumes negative values, tending to
$-p_R$. Parameters: $p_L=0.23$, $p_R=0.35$, $k_B
T=200$ $\mu$eV, $\G_0=10$ $\mu$eV, $\e_0=0.5$ meV and U=1
meV.}\label{fig4}
\end{figure}

\emph{Spin-flip effects}. In figure (\ref{fig4}) we show the
current polarization against bias voltage for distinct spin-flip parameter $R$. The polarization is calculated for the left and right
leads, according to the formula
$\wp^\eta=(I^\uparrow_\eta-I^\downarrow_\eta)/(I^\uparrow_\eta+I^\downarrow_\eta)$,
where $\eta=L,R$. For $R=0$ (solid line) we have $\wp^L=\wp^R$ for
all biases. This curve is the same as seen in Fig. \ref{fig3}(d).
When $R \neq 0$ these two polarizations depart from each
other. The $\wp^L$ increases with $R$ tending to reach the left
lead polarization value $p_L=0.23$. In contrast, $\wp^R$ assume
negative values tending to $-p_R$ as $R$ increases. This shows that even though we have a constant total current along
the system ($I_L^\uparrow+I_L^\downarrow=I_R^\uparrow+I_R^\downarrow$), its
polarization can change across the system when $R \neq 0$, i.e., when there is  a transverse magnetic field applied on 
the dot which coherently rotates the spin.

\begin{figure}[tbp]
\par
\begin{center}
\epsfig{file=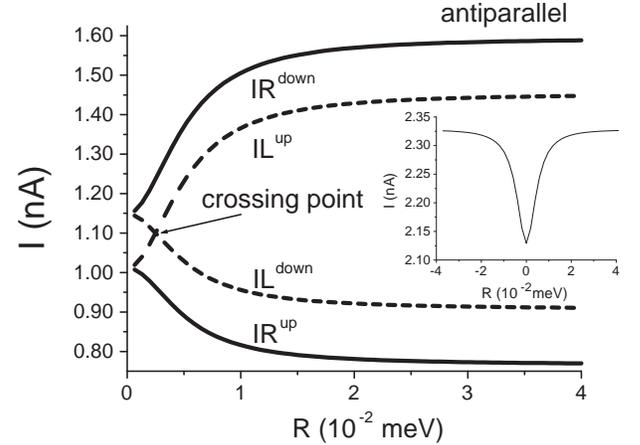, width=0.45\textwidth}
\end{center}
\caption{Spin-resolved currents versus $R$. For $R=0$
we have $I_L^\uparrow=I_R^\uparrow$ and
$I_L^\downarrow=I_R^\downarrow$ with $I_\eta^\uparrow$ and
$I_\eta^\downarrow$ being different from each other. For
increasing $R$, $I_L^\uparrow$ and $I_R^\downarrow$ grow while
$I_L^\downarrow$ and $I_R^\uparrow$ decreases. For a particular
$R$ there is a crossing between $I_L^\uparrow$ and
$I_L^\downarrow$, thus resulting simultaneously in a non-polarized
left-current and a polarized right-current. Parameters:
$p_L=0.23$, $p_R=0.35$, $k_B T=200$ $\mu$eV, $\G_0=10$ $\mu$eV,
$\e_0=0.5$ meV, U=1 meV and bias=4 meV.}\label{fig5}
\end{figure}

Figure \ref{fig5} shows $I_\eta^\s$ against $R$. For $R=0$,
$I^\uparrow_L=I^\uparrow_R$ and $I_L^\downarrow=I_R^\downarrow$ as
expected. For $R \neq 0$ these equalities disappear, with
$I_L^\uparrow$ and $I_R^\downarrow$ increasing and
$I_L^\downarrow$ and $I_R^\uparrow$ decreasing with $R$. This
leads to an enhancement of $\wp^L$ and $|\wp_R|$ as seen in Fig.
\ref{fig4}. Interestingly, there is a crossing point between
$I_L^\uparrow$ and $I_L^\downarrow$ around $R \approx \G_0/4$. So
for this particular $R$ the total current becomes unpolarized in
the emitter (left) lead and relatively high polarized in the
collector (right) lead. This means that it is possible to change
the current polarization from emitter to collector lead by precessing
the electron spin in the quantum dot.

In the inset of Fig. \ref{fig5} we show the total current against $R$. 
This curve resembles a typical Hanle resonance.\cite{hanle,mb05}
Similarly to Ref. [\onlinecite{mb05}], here we can say that in the AP configuration and positive bias (i.e., with left being the emitter)
the dot tends to be more up populated due to the majority up population in the emitter and the majority down population in the collector lead.
On average a transverse magnetic field tends to increase the spin down component in the dot along the down magnetization
of the collector lead. As a result the electron can more easily tunnel into the right ferromagnet and the current increases.

\subsection{Shot Noise}

\begin{figure}[t]
\par
\begin{center}
\epsfig{file=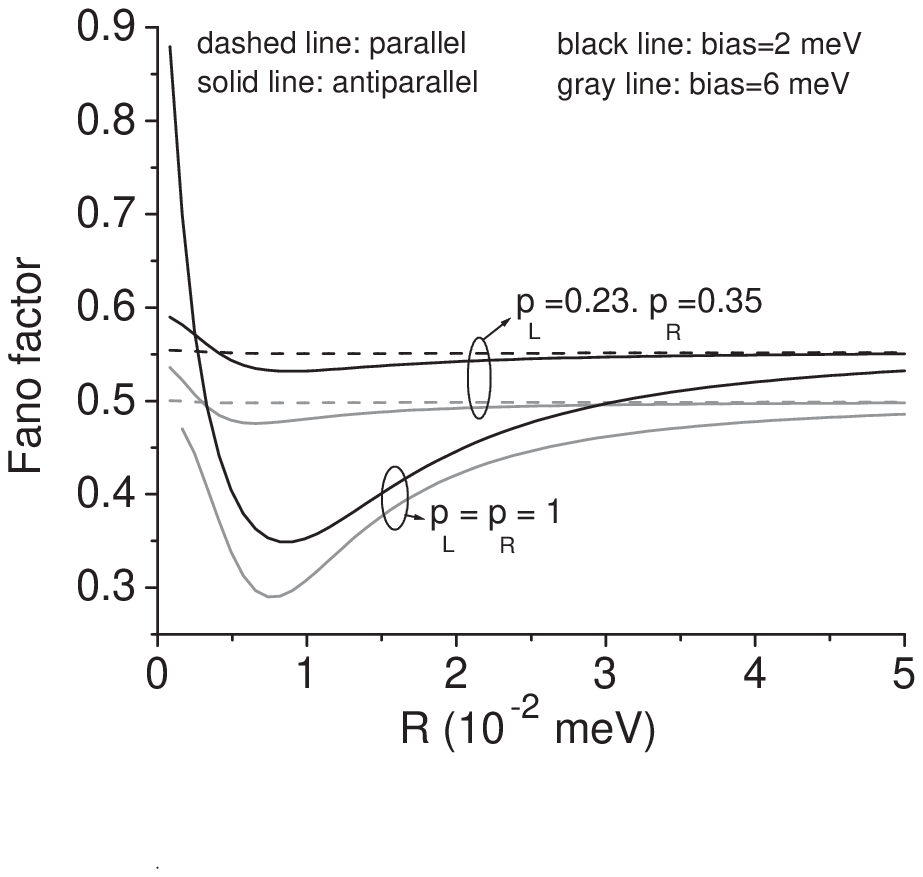, width=0.45\textwidth}
\end{center}
\caption{Fano factor against $R$ for two fixed bias voltages corresponding to single (bias = 2 meV)
and double occupancy (bias = 6 meV) in both P and AP alignments. In the P case the Fano factor is
approximatelly constante. In contrast, in the AP alignment the Fano factor can be suppressed
due to spin-flip, reaching values below 0.5. In particular, for fully polarized leads ($p_L=p_R=1$) the
Fano factor can be strongly suppressed, assuming values close to 0.3 when double occupancy is allowed. Parameters: $\e_0=0.5$ meV, $U=1$ meV, $\G_0=10$ $\mu$eV and $k_B T=200$
$\mu$eV.}\label{fig6}
\end{figure}

Figure \ref{fig6} shows the Fano factor, $\g=S_{LL}/2eI_L$,
against $R$ in the AP configurations. The P alignment gives approximately
insensitive Fano factor with respect to $R$. In the AP
configuration, the Fano factor can be suppressed with $R$,
reaching values below 0.5. This suppression can be further
intensified by increasing the lead polarization parameters $p_L$
and $p_R$. In particular, for fully spin-polarized leads ($p=1$) AP aligned,
the Fano factor reaches values close to 0.3 when double occupancy is allowed (bias = 6 meV) and it attains
0.35 in the single occupancy regime (bias = 2 meV).\cite{fgb02} For fully polarized leads in the P configuration
the Fano factor remains at 0.5 independently of $R$. 

A simple physical picture for this additional suppression of $\g$ is as follows. Consider an up spin sitting on the dot. A second up spin trying to hop onto it is Pauli blocked, till the first electron tunnel to the collector lead or undergo a coherent spin-flip. If the spin-flip is fast enough (faster than the into/out tunneling processes) the first electron can return to its original state (via another spin-flip), instead of tunneling out of the dot. This blocks additionally the second up spin, consequently suppressing even further the noise.

\begin{figure}[h]
\par
\begin{center}
\epsfig{file=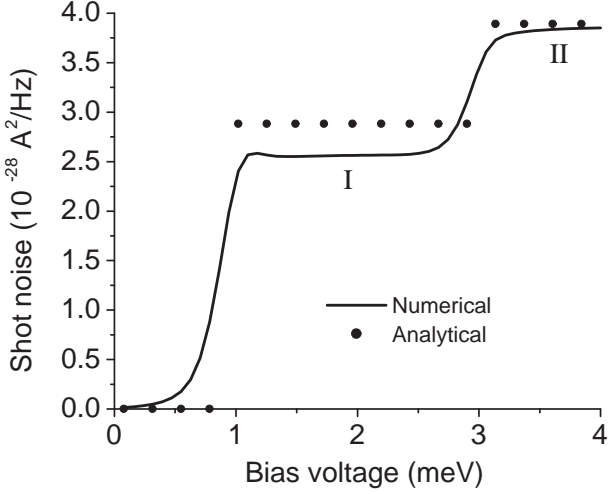, width=0.45\textwidth}
\end{center}
\caption{Comparison of the shot noise for the present approach (numerical) and
the one derived in Ref. [\onlinecite{at03}] (analytical). While the second
plateau (II) coincides in both approaches, the first plateau (I) differs. We
attribute this contrast to the Hartree-Fock type approximation used in
our calculation. Parameters: $p_L=p_R=0$, $k_B T=50$ $\mu$eV,
$\G_0=10$ $\mu$eV , $R=0$, $\e_0=0.5$ meV and $U=1$
meV.}\label{fig7}
\end{figure}

We note that for any $R$ when we go from the single (bias = 2 meV) to the double (bias = 6 meV) occupation regimes a reduction of the Fano factor is observed in both P and AP alignments (cf. solid black to solid gray lines and dashed black to dashed gray lines). This general feature was already predicted in Ref. [\onlinecite{mb06}], where a diagrammatic formulation for the noise is derived. It is valid to mention that in the present study we have performed an S-matrix expansion (Appendix B) to the noise which could in principle be mapped into a Feynman diagrammatic formulation. Comparing our results with previous findings in the literature we observe a difference between them in the single-occupancy regime.\cite{singleocup} Figure \ref{fig7} shows a comparison for the shot noise obtained from Eq. (\ref{SLL}) and from analytical results in Ref. [\onlinecite{at03}] (derived for $p_L=p_R=0$ and $R=0$). While the second plateau [II in Fig. (\ref{fig7})] coincides in both
numerical and analytical cases, the first plateaus (I in the plot) do not coincide.\cite{comparison} 
This disagreement is related to the Hartree-Fock factorization underlying our calculation.

It is yet valid to note that without Coulomb interaction
($U=0$) and for fully spin polarized leads antiparallel aligned, the Fano factor is given by
\begin{equation}\label{fano_similar_mulburn}
\gamma=1-\frac{1}{2} \frac{\beta^2 (5+\beta^2)}{(1+\beta^2)^2},
\end{equation}
where $\beta=R/\Gamma_0$. Eq. (\ref{fano_similar_mulburn}) is also found for a three tunneling
barriers junction.\cite{hbs97} Hence, for fully spin-polarized AP
leads and non-vanishing spin-flip, the FM-QD-FM setup resembles a
three-barrier geometry.

\section{CONCLUSION}

Using the nonequilibrium Green-function technique we have studied
the transport properties of a quantum dot coupled to two
ferromagnetic leads. We consider both parallel (P) and
antiparallel (AP) alignments of the lead polarizations. Coulomb
interaction and coherent spin-flip are included in our model. We
find that for distinct ferromagnetic leads the interplay between
Coulomb interaction and spin accumulation translates into an
enhancement and a suppression of the current polarization $\wp$ in
AP and P cases, respectively, depending on the bias. We also
observe that the spin-flip can change the current polarization
when it flows from the emitter to the collector. It is even
possible to have a polarized current in the collector while it is
unpolarized in the emitter. We have derived an expression for 
the noise [Eq. (\ref{SLL})] which exactly accounts for spin-flip but only approximately for Coulomb interaction. 
Finally, we found a suppression of the Fano factor to values well below 1/2 due to spin-flip.

The authors acknowledge support for this work from the Brazilian agencies CNPq, CAPES, FAPESP and IBEM and 
the FiDiPro program of the Finnish Academy. The authors thank J. K\"onig and J. Martinek for helpful discussions.

\appendix

\section{Dot Green Functions}

Here we present in some detail the calculation of the dot Green
functions $ \mathbf{G}^r$, $\mathbf{G}^a$, $\mathbf{G}^>$ and
$\mathbf{G}^<$, used in the current and noise expressions. The
starting point is to derive an equation of motion for the
contour-ordered Green function $G_{\sigma\sigma^{\prime}}(
\tau,\tau^{\prime})=-i\langle T_c d_\s(\tau)
d_{\sigma^{\prime}}^\dagger(\tau^{\prime})\rangle$, and then, via
analytical continuation rules, to determine these Green functions.
After a straightforward calculation via equation of motion we find
\begin{eqnarray}  \label{G}
\mathbf{G}(\tau,\tau^{\prime})&=&\mathbf{G}^0(\tau,\tau^{\prime})
\nonumber
\\
&+&\int \int d\tau_1 d\tau_2 \mathbf{G}^0(\tau,\tau_1) \mathbf{\Sigma}%
(\tau_1,\tau_2) \mathbf{G}(\tau_2,\tau^{\prime})  \nonumber \\
&+&\int d\tau_1 \mathbf{G}^{0}(\tau,\tau_1) U \mathbf{G}^{(2)}(\tau_1,\tau^{%
\prime}),
\end{eqnarray}
where the components of $\mathbf{G}^{(2)}$ and $\mathbf{\Sigma}$ are $%
G_{ij}^{(2)}(\tau,\tau^{\prime})=-i\langle T_c
n_{\overline{i}}(\tau)
d_i(\tau) d_{j}^\dagger(\tau^{\prime})\rangle$, and $\Sigma_{ij}(\tau,\tau^{%
\prime})=\sum_{k\eta\sigma}
\frac{1}{2}(-1)^{(i+j)\delta_{\sigma\downarrow}}
|t_{k\sigma}|^2 g_{k\sigma\eta}(\tau,\tau^{\prime})$, respectively. $\mathbf{%
G}^0$ is the dot Green function without both the coupling to leads and
the Coulomb interaction.

Following the equation of motion expansion we find for $\mathbf{G}^{(2)}(\tau
_{1},\tau ^{\prime })$,
\begin{eqnarray}\label{G2APbefdy}
\mathbf{G}^{(2)}(\tau _{1},\tau ^{\prime })
&=&\mathbf{G}^{0(2)}(\tau _{1},\tau ^{\prime })+\int \int d\tau
_{2}d\tau _{3}\mathbf{G}^{0(2)}(\tau
_{1},\tau _{2})  \nonumber  \label{G2} \\
&\times &\mathbf{\Sigma }(\tau _{2},\tau _{3})\mathbf{G}(\tau
_{3},\tau ^{\prime }),
\end{eqnarray}%
where $\mathbf{G}^{0(2)}(\tau _{1},\tau ^{\prime })$ satisfies the identity
\begin{eqnarray}
(i\frac{\partial}{\partial \t}-\e_i-U)G_{ij}^{0(2)}(\t,\t')&=&\d_{ij}\d(t-t') \langle n_{\overline{i}} \rangle - \nonumber \\ && \d_{\overline{i}j} \d(t-t') \langle d_{\overline{i}}^\dagger d_i \rangle.
\end{eqnarray}
In Eq. (\ref{G2APbefdy}) we have used the following approximations,\cite{standard_approx}
\begin{eqnarray}
\langle Td_{\overline{i}}(t)d_{i}(t)c_{k\sigma \eta }^{\dagger
}(t)d_{j}^{\dagger }(t^{\prime })\rangle &=&0  \nonumber \\
\langle Td_{\overline{i}}^{\dagger }(t)d_{i}(t)c_{k\sigma \eta
}(t)d_{j}^{\dagger }(t^{\prime })\rangle &=&\langle d_{\overline{i}%
}^{\dagger }d_{i}\rangle \langle Tc_{k\sigma \eta
}(t)d_{j}^{\dagger
}(t^{\prime })\rangle  \nonumber \\
\langle Td_{\overline{i}}^{\dagger
}(t)d_{\overline{i}}(t)c_{k\sigma \eta
}(t)d_{j}^{\dagger }(t^{\prime })\rangle &=&\langle d_{\overline{i}%
}^{\dagger }d_{\overline{i}}\rangle \langle Tc_{k\sigma \eta
}(t)d_{j}^{\dagger }(t^{\prime })\rangle.  \nonumber
\end{eqnarray}
Observe that Eq.(\ref{G2}) closes the system of equations (\ref{G})-(\ref{G2}). Substituting (\ref{G2}) into (\ref{G}) we find a Dyson
equation for $\mathbf{G}$
\begin{eqnarray}
\mathbf{G}(\tau ,\tau ^{\prime })
&=&\widetilde{\mathbf{G}}^{0}(\tau ,\tau ^{\prime })+\int \int
d\tau _{1}d\tau _{2}\widetilde{\mathbf{G}}^{0}(\tau
,\tau _{1})  \nonumber  \label{Gbefore} \\
&\times &\mathbf{\Sigma }(\tau _{1},\tau _{2})\mathbf{G}(\tau
_{2},\tau ^{\prime }),
\end{eqnarray}%
where
\begin{equation}
\widetilde{\mathbf{G}}^{0}(\tau ,\tau ^{\prime
})=\mathbf{G}^{0}(\tau ,\tau
^{\prime })+\int d\tau _{1}\mathbf{G}^{0}(\tau ,\tau _{1})U\mathbf{G}%
^{0(2)}(\tau _{1},\tau ^{\prime }).  \label{G0before}
\end{equation}
Applying analytical continuation rules in Eqs. (\ref{Gbefore}) and
(\ref{G0before}) we find
\begin{eqnarray}\label{GRAppendix}
\mathbf{G}^{r}(t,t^{\prime })
&=&\widetilde{\mathbf{G}}^{0r}(t,t^{\prime
})+\int \int dt_{1}dt_{2}\widetilde{\mathbf{G}}^{0r}(t,t_{1})  \nonumber \\
&\times &\mathbf{\Sigma
}^{r}(t_{1},t_{2})\mathbf{G}^{r}(t_{2},t^{\prime }),
\end{eqnarray}
\begin{equation}\label{G0RAppendix}
\widetilde{\mathbf{G}}^{0r}(t,t^{\prime
})=\mathbf{G}^{0r}(t,t^{\prime })+\int
dt_{1}\mathbf{G}^{0r}(t,t_{1})U\mathbf{G}^{0(2)}(t_{1},t^{\prime
}),
\end{equation}
and also the Keldysh equation
\begin{equation}\label{GlesserAppendix}
\mathbf{G}^{<}(t,t^{\prime })=\int \int dt_{1}dt_{2}\mathbf{G}^{r}(t,t_{1})%
\mathbf{\Sigma }^{<}(t_{1},t_{2})\mathbf{G}^{a}(t_{2},t^{\prime
}).
\end{equation}
Via Fourier transform of Eqs. (\ref{GRAppendix}) and
(\ref{G0RAppendix}) we obtain
\begin{equation}\label{A9}
\mathbf{G}^{r}(\e)=[\widetilde{\mathbf{G}}^{0r^{-1}}(\e)-\mathbf{
\Sigma }^{r}(\e)]^{-1},
\end{equation}
\begin{equation}\label{A10}
\widetilde{\mathbf{G}}^{0r}(\e)=\mathbf{G}^{0r}(\e)+\mathbf{G}
^{0r}(\e)U\mathbf{G}^{0(2)}(\e),
\end{equation}%
and of Eq. (\ref{GlesserAppendix}) we find
\begin{equation}\label{A11}
\mathbf{G}^{<}(\e)=\mathbf{G}^{r}(\e)\mathbf{\Sigma }^{<}(\e)%
\mathbf{G}^{a}(\e),
\end{equation}%
where
\begin{equation}\label{A12}
\mathbf{G}^{0r}(\e)=\left(
\begin{array}{cc}
\frac{1}{\e -\epsilon _{1}+i0^+ } & 0 \\
0 & \frac{1}{\e -\epsilon _{2}+i0^+ }%
\end{array}%
\right)
\end{equation}%
and
\begin{equation}
\mathbf{G}^{0(2)r}(\e)=\left(
\begin{array}{cc}
\frac{\langle n_{2}\rangle }{\e -\epsilon _{1}-U+i0^+ } & -\frac{%
\langle d_{2}^{\dagger }d_{1}\rangle }{\e -\epsilon _{1}-U+i0^+ } \\
-\frac{\langle d_{1}^{\dagger }d_{2}\rangle }{\e -\epsilon
_{2}-U+i0^+ }
& \frac{\langle n_{1}\rangle }{\e -\epsilon _{2}-U+i0^+ }%
\end{array}%
\right) .
\end{equation}%
The retarded self energy is given by
\begin{equation}
\Sigma ^{r}=-\frac{i}{2}\left(
\begin{array}{cc}
\Gamma _{\uparrow }+\Gamma _{\downarrow } & \Gamma _{\uparrow
}-\Gamma
_{\downarrow } \\
\Gamma _{\uparrow }-\Gamma _{\downarrow } & \Gamma _{\uparrow
}+\Gamma
_{\downarrow }%
\end{array}%
\right) ,
\end{equation}%
and the lesser self energy is defined as
\begin{equation}\label{A15}
\Sigma _{lm}^{<}=\sum_{\s}\frac{i}{2}(-1)^{(l+m)\delta _{\sigma
\downarrow }}(n_{L}\Gamma _{\s}^{L}+n_{R}\Gamma _{\s}^{R}).
\end{equation}

\section{S-matrix expansion for the noise}

According to Eq. (\ref{Setaeta'middle}) the noise is given in
terms of the four-operator Green functions $g^{(2)}_i
(\tau,\tau^{\prime})$ [$i=1,2,3,4$]. To determine their equations of motion we
develop an $S$-matrix expansion as we illustrate below for
$g^{(2)}_1(\tau,\tau^{\prime})$. The first step is to transform the operators from the Heisenberg to the interaction picture,
\begin{eqnarray}  \label{g1appendix}
g^{(2)}_1 (\tau,\tau^{\prime})&=&i^2\langle T_c S \widetilde{c}%
_{k\sigma\eta}^{\dagger}(\tau)\widetilde{d}_i(\tau)\widetilde{c}%
_{k^{\prime}\sigma^{\prime}\eta^{\prime}}^{\dagger}(\tau^{\prime})\widetilde{%
d}_{j}(\tau^{\prime})\rangle,\nonumber
\end{eqnarray}
where the tilde denotes the operators in the interaction picture, i.e.,
\begin{equation}
d_i(t)=v^\dagger (t,t_0) \widetilde{d}_i(t) v(t,t_0),
\end{equation}
with
\begin{equation}
v(t,t_0)=T e^{-i \int_{t_0}^t dt' \widetilde{H}_T (t')},
\end{equation}
and a similar definition for the $c$ operator. The operator $T$ is the time-ordering operator. The
$S$-matrix in $g^{(2)}_1 (\tau,\tau^{\prime})$ is defined as
\begin{equation}
S=T_c e^{-i\int_c d\tau_1 \widetilde{H}_T (\tau_1)}.  \label{B2}
\end{equation}
Expanding $S$ we find
\begin{eqnarray}  \label{appendixg1}
&& g^{(2)}_1(\tau,\tau^{\prime})= i^2 \sum_{n=0}^{\infty}
\frac{(-i)^{n}}{n!}
\nonumber \\
&&  \phantom {xx} \times \langle T_c
\widetilde{c}_{k\sigma\eta}^{\dagger}(\tau)\widetilde{d}
_i(\tau)\widetilde{c}_{k^{\prime}\sigma^{\prime}\eta^{\prime}}^{\dagger}(%
\tau^{\prime})\widetilde{d}_{j}(\tau^{\prime})  \nonumber \\
&& \times [\int d\tau_1 \sum_{k_1 \sigma_1 \eta_1 i_1} \frac{(-1)^{i_1\delta_{%
\sigma_1\downarrow}}}{\sqrt{2}} (t_{k_1 \sigma_1}
\widetilde{c}_{k_1
\sigma_1 \eta_1}^\dagger (\tau_1) \widetilde{d}_{i_1} (\tau_1)  \nonumber \\
&&  \phantom{xxxxxxx} +t_{k_1
\sigma_1}^*\widetilde{d}_{i_1}^\dagger(\tau_1) \widetilde{c}_{k_1
\sigma_1 \eta_1} (\tau_1) ]^{n}\rangle, \label{B3}
\end{eqnarray}
where the lowest-order nonzero term in the expansion is that of
$n=2$. Since we assume noninteracting leads, we can factorize the angle
bracket in Eq. (\ref{B3}) into a product of the lead and dot parts.
We then apply Wick's theorem to the lead part. This results in
\begin{eqnarray}  \label{g1intheappendix}
g^{(2)}_1(\tau,\tau^{\prime})&=&t_{k\sigma}^*
t_{k^{\prime}\sigma^{\prime}}^* \int \int d\tau_1 d\tau_2
\sum_{k_1 \sigma_1
\eta_1} \sum_{k_2 \sigma_2 \eta_2}  \nonumber \\
&&\sum_{i_1 i_2=1,2} \frac{1}{2}(-1)^{i_1 \delta_{\sigma_1
\downarrow}}
(-1)^{i_2 \delta_{\sigma_2 \downarrow}}  \nonumber \\
&&(-i)\langle T_c \widetilde{c}_{k_1\sigma_1\eta_1}(\tau_1) \widetilde{c}%
_{k\sigma\eta}^\dagger (\tau)\rangle  \nonumber \\
&& (-i)\langle T_c \widetilde{c}_{k_2\sigma_2\eta_2}(\tau_2) \widetilde{c}%
_{k^{\prime}\sigma^{\prime}\eta^{\prime}}^\dagger (t^{\prime})
\rangle
\nonumber \\
&& \langle T_c S \widetilde{d}_i (\tau)
\widetilde{d}_{j}(\tau^{\prime}) \widetilde{d}_{i_1}^\dagger
(\tau_1)\widetilde{d}_{i_2}^\dagger(\tau_2) \rangle.
\end{eqnarray}
In Eq. (\ref{g1intheappendix}) we have contracted $\widetilde{c}%
_{k_1\sigma_1\eta_1}$ with $\widetilde{c}_{k\sigma\eta}^\dagger$, and $%
\widetilde{c}_{k_2\sigma_2\eta_2}$ with $\widetilde{c}^\dagger_{k^{\prime}%
\sigma^{\prime}\eta^{\prime}}$. This is one choice among $n(n-1)$
possible contractions. Since all of them yields the same result, we
simply multiply the chosen pairing by $n(n-1)$. This factor
cancels part of the factorial $n!$ in Eq.(\ref{B3}), thus resulting
in the $S$-matrix in the last angle bracket of Eq. (\ref
{g1intheappendix}).

The first and second averages in (\ref{g1intheappendix}) give $%
\delta_{k\sigma\eta,k_1\sigma_1\eta_1}$ and $\delta_{k^{\prime}\sigma^{%
\prime}\eta^{\prime},k_2,\sigma_2,\eta_2}$, respectively, so the sums over $%
(k_1,\sigma_1,\eta_1)$ and $(k_2,\sigma_2,\eta_2)$ disappear. Defining $%
g_{k\sigma\eta}(\tau_1,\tau)=-i\langle T_c \widetilde{c}_{k\sigma\eta}(%
\tau_1) c_{k\sigma\eta}^\dagger (\tau)\rangle$ and $g_{k^{\prime}\sigma^{%
\prime}\eta^{\prime}}(\tau^{\prime},\tau_2)=-i\langle T_c \widetilde{c}%
_{k^{\prime}\sigma^{\prime}\eta^{\prime}}(\tau^{\prime}) \widetilde{c}%
_{k^{\prime}\sigma^{\prime}\eta^{\prime}}^\dagger
(\tau_2)\rangle$, we can rewrite (\ref{g1intheappendix}) as
\begin{eqnarray}
&& g^{(2)}_1(\tau,\tau^{\prime})=t_{k\sigma}^*
t_{k^{\prime}\sigma^{\prime}}^* \int \int d\tau_1 d\tau_2  \nonumber \\
&\times & \sum_{i_1,i_2=1,2} \frac{1}{2}(-1)^{i_1 \delta_{\sigma_1
\downarrow}} (-1)^{i_2 \delta_{\sigma_2 \downarrow}}  \nonumber \\
&\times & g_{k\sigma\eta}(\tau_1,\tau)
g_{k^{\prime}\sigma^{\prime}\eta^{\prime}}(\tau_2,\tau^{\prime})
\langle T_c d_i(\tau) d_{j}(\tau^{\prime}) d_{i_1}^\dagger (\tau_1)
d_{i_2}^\dagger (\tau_2) \rangle.  \label{g21_before_wick}\nonumber
\end{eqnarray}

For the $U=0$ case the calculation is straightforward. By applying Wick's theorem in the
four operators Green function, we find
\begin{eqnarray}  \label{g1finalappendix}
&&g^{(2)}_1(\tau,\tau^{\prime})=t_{k\sigma}^*
t_{k^{\prime}\sigma^{\prime}}^* \int \int d\tau_1 d\tau_2
\nonumber\\ && \phantom{x} \times \sum_{i_1 i_2} \frac{1}{2}
(-1)^{i_1 \delta_{\sigma\downarrow}} (-1)^{i_2
\delta_{\sigma^{\prime}\downarrow}} g_{k\sigma\eta}(\tau_1,\tau)
g_{k^{\prime}\sigma^{\prime}\eta^{\prime}}(\tau_2,\tau^{\prime})
\nonumber\\ && \phantom{x} \times \{G_{i i_1}(\tau,\tau_1) G_{j
i_2}(\tau^{\prime},\tau_2) -G_{i i_2}(\tau,\tau_2)G_{j
i_1}(\tau^{\prime},\tau_1)\}\nonumber\\
\end{eqnarray}
where $G_{i i_1}(\tau,\tau_1)=-i\langle T_c d_i(\tau)
d_{i_1}^\dagger (\tau_1) \rangle$, plus analogous definitions for
the other Green functions. A similar calculation yields Eq.
(\ref{g2}) for $g^{(2)}_2(\tau,\tau^{\prime})$. In the presence of
the Coulomb interaction ($U \neq 0$) Eq. (\ref{g1finalappendix})
is no longer exact and the full diagrammatic expansion should be
considered in order to find an accurate noise expression. However, this is
a formidable task since it involves not only the usual many body expansion
but also the analytical continuation of two and more particles Green functions.
So as a first approximation we use Eq. (\ref{g1finalappendix}) even in the presence of
the Coulomb interaction.

\section{``Linked-Cluster Theorem'' for the noise expansion}

Equations (\ref{g1})-(\ref{g4}) are composed of what we call connected and disconnected terms. Here we show
that the disconnected parts cancel identically the term $2I_\eta^2$ in Eq. (\ref{Setaeta'middle}). Writing
explicitly the disconnected term of Eq.(\ref{g1}), we have
\begin{eqnarray}
&&g_{1 \rm{disc}}^{(2)}(\tau ,\tau ^{\prime })=t_{k\sigma }^{\ast
}t_{k^{\prime
}\sigma^{\prime }}^{\ast }  \sum_{i_1 i_2} \frac{1}{2} (-1)^{i_1\d_{\s \downarrow}} (-1)^{i_2 \d_{\s^\prime \downarrow}}\nonumber  \label{g1disc} \\
&& \phantom{xxxxxx} \times \int d\tau _{1}G_{i i_1}(\tau ,\tau _{1})g_{k\sigma \eta
}(\tau
_{1},\tau^+ )  \nonumber \\
&& \phantom{xxxxxx} \times \int d\tau _{2}G_{j i_2}(\tau ^{\prime },\tau
_{2})g_{k^{\prime }\sigma ^{\prime }\eta ^{\prime }}(\tau
_{2},\tau ^{\prime +}),
\end{eqnarray}
where the $+$ sign on one of the $\t$ and $\t'$ is just reminder that the sequence of operators $c_{k \s \eta}^\dagger (\t) d_i(\t)$ and
$c_{k' \s' \eta'}^\dagger (\t') d_j (\t')$ in the main definition of $g_1^{(2)}(\t,\t')$ (beginning of Sec. III-D) should be
preserved during the following calculation. Applying the analytic continuation rules we obtain
\begin{eqnarray}\label{g1discles}
g_{1 \rm{disc}}^{(2)>}(t,t^{\prime }) &=&t_{k\sigma }^{\ast }t_{k^{\prime }\sigma
^{\prime }}^{\ast }\nonumber \\
&&F_{i \s,k\eta \sigma }(t,t)F_{j \s',k^{\prime }\eta ^{\prime
}\sigma ^{\prime }}(t^{\prime },t^{\prime }),
\end{eqnarray}%
with
\begin{eqnarray}\label{Fdef1}
&&F_{i \s,k\eta \sigma }(t,t) =\sum_{i_1} \frac{1}{\sqrt{2}} (-1)^{i_1 \d_{\s \downarrow}} \int
dt_{1} \times \nonumber \\ && \phantom{x} [G_{i i_1}^{r}(t,t_{1})g_{k\sigma
\eta }^{<}(t_{1},t)  +G_{i i_1}^{<}(t,t_{1})g_{k\sigma \eta }^{a}(t_{1},t)],
\end{eqnarray}%
and a similar definition for $F_{j\s',k^{\prime }\eta ^{\prime
}\sigma ^{\prime }}(t^{\prime },t^{\prime })$. Similarly, from
Eq. (\ref{g2}), we have
\begin{eqnarray}\label{g2disc} 
&&g^{(2)}_{2 \rm{disc}}(\tau,\tau^{\prime})=t_{k\sigma}^* t_{k^{\prime}\sigma^{\prime}} \sum_{i_1,i_2=1,2} \frac{1}{2} (-1)^{i_1
\delta_{\sigma\downarrow}} (-1)^{i_2 \delta_{\sigma^{\prime}\downarrow}} \nonumber \\  
&&  \phantom{xxx} \times \int  d\tau_1  G_{i i_1}(\tau,\tau_1) g_{k\sigma\eta}(\tau_1,\tau^+)
\nonumber \\ && \phantom{xxx} \times \int d\tau_2 g_{k^{\prime}\sigma^{\prime}\eta^{\prime}}(\tau^{\prime},\tau_2) G_{i_2 j}(\tau_2,\tau^{\prime +}).
\end{eqnarray}
which, after analytic continuation, can be expressed as
\begin{eqnarray}\label{g2discless}
g_{2 \rm{disc}}^{(2)>}(t,t^{\prime }) &=&-t_{k\sigma }^{\ast
}t_{k^{\prime }\sigma ^{\prime }}\nonumber\\
&&F_{i \s,k \eta \sigma }(t,t)F_{j \s',k^{\prime }\eta ^{\prime
}\sigma ^{\prime }}^{\ast }(t^{\prime },t^{\prime }).
\end{eqnarray}\label
Using the identities Eqs. (\ref{g3})-(\ref{g4}) we obtain
\begin{eqnarray}\label{g3discles}
g_{3 \rm{disc}}^{(2)>}(t,t^{\prime }) &=&-t_{k\sigma }t_{k^{\prime
}\sigma ^{\prime }}^{\ast }\nonumber
\label{g3disc} \\
&&F_{i \s, k \eta \s }^{\ast }(t,t)F_{j \s',k'
\eta'\s'}(t^{\prime },t^{\prime }),
\end{eqnarray}%
and
\begin{eqnarray}\label{g4discles}
g_{4 \rm{disc}}^{(2)>}(t,t^{\prime }) &=&t_{k\sigma }t_{k^{\prime
}\sigma ^{\prime }} \nonumber
\label{g4disc} \\
&&F_{i \s,k \eta \s }^{\ast }(t,t)F_{j \s',k' \eta'\s'}^{\ast
}(t^{\prime },t^{\prime }).
\end{eqnarray}%
From Eqs. (\ref{Sdefinition}) and (\ref{Setaeta'middle}) we note
that
\begin{eqnarray}
\langle \{\widehat{I}_{\eta }(t),\widehat{I}_{\eta ^{\prime }}(t^{\prime
})\}\rangle _{\rm{disc}} &=&e^{2}\sum_{kk^{\prime }\sigma \sigma ^{\prime } i j} \frac{1}{2}
(-1)^{i\delta_{
\sigma\downarrow}}(-1)^{j\delta_{\sigma^{\prime}\downarrow}} 
\nonumber  \label{IetaIeta'disc} \\
&&\{t_{k\sigma }t_{k^{\prime }\sigma ^{\prime }}g_{1 \rm{disc}}^{(2)>}(t,t^{\prime
})  \nonumber \\
&&-t_{k\sigma }t_{k^{\prime }\sigma ^{\prime }}^{\ast
}g_{2 \rm{disc}}^{(2)>}(t,t^{\prime })  \nonumber \\
&&-t_{k\sigma }^{\ast }t_{k^{\prime }\sigma ^{\prime
}}g_{3 \rm{disc}}^{(2)>}(t,t^{\prime })  \nonumber \\
&&+t_{k\sigma }^{\ast }t_{k^{\prime }\sigma ^{\prime }}^{\ast
}g_{4 \rm{disc}}^{(2)>}(t,t^{\prime })\}  \nonumber \\
&&+h.c.
\end{eqnarray}%
Using Eqs. (\ref{g1disc}), (\ref{g2disc}), (\ref{g3disc}) and (\ref{g4disc}) in Eq. (\ref{IetaIeta'disc}) we find
\begin{eqnarray}\label{IIdisc}
&&\langle \{\widehat{I}_{\eta }(t),\widehat{I}_{\eta ^{\prime
}}(t^{\prime })\}\rangle _{\rm{disc}}=2e^{2}\sum_{k k' \s \s' i j}
|t_{k\sigma}|^{2}|t_{k^{\prime }\sigma ^{\prime }}|^{2}  \nonumber \\ &&\frac{1}{2}
(-1)^{i\delta_{\sigma\downarrow}}(-1)^{j\delta_{\sigma^{\prime}\downarrow}} [F_{i \s,k \eta \s}(t,t)+F_{i \s,k \eta \s }^{\ast
}(t,t)] \nonumber \\
&& \phantom{xxx} \times[F_{j \s',k^{\prime }\eta ^{\prime }\sigma ^{\prime
}}(t^{\prime },t^{\prime })+F_{j \s',k^{\prime }\eta ^{\prime
}\sigma ^{\prime }}^{\ast }(t^{\prime },t^{\prime })].
\end{eqnarray}%
On the other hand we can write the current as
\begin{eqnarray}\label{IetaF}
\langle \widehat{I}_{\eta }(t)\rangle
&=&2 e \mathrm{Re} \sum_{k\sigma i}|t_{k\sigma}|^{2} \frac{(-1)^{i \d_{\s \downarrow}}}{\sqrt{2}} F_{i\s,k \eta \s}(t,t).\nonumber\\ &=& e \sum_{k\sigma i}|t_{k\sigma}|^{2} \frac{(-1)^{i \d_{\s \downarrow}}}{\sqrt{2}} \nonumber \\ && \phantom{xxx} \times[ F_{i\s,k \eta \s}(t,t)+ F_{i\s,k \eta \s}^{\ast}(t,t)] 
\end{eqnarray}%
Squaring Eq. (\ref{IetaF}) and multiplying it by two, we find
\begin{eqnarray}\label{2IetaIeta'}
&&2\langle \widehat{I}_{\eta }(t)\rangle \langle \widehat{I}_{\eta
^{\prime }}(t^{\prime })\rangle
=2e^{2}\sum_{k k' \sigma \sigma' i j} |t_{k\sigma
}|^{2}|t_{k^{\prime }\sigma ^{\prime }}|^{2}  \nonumber \\
&& \times \frac{1}{2}
(-1)^{i\delta_{\sigma\downarrow}}(-1)^{j\delta_{\sigma^{\prime}\downarrow}} [F_{i \s,k \eta \s}(t,t)+F_{i \s,k \eta \s}^{\ast }(t,t)]
\nonumber \\
&& \phantom{xxxx} \times [F_{j \s',k^{\prime }\eta ^{\prime }\sigma ^{\prime
}}(t^{\prime },t^{\prime })+F_{j \s',k^{\prime }\eta ^{\prime
}\sigma ^{\prime }}^{\ast }(t^{\prime },t^{\prime })].
\end{eqnarray}
Hence, Eq. (\ref{2IetaIeta'}) cancels identically with (\ref{IIdisc}), i.e.,
\begin{equation}
\langle \{\widehat{I}_{\eta }(t),\widehat{I}_{\eta ^{\prime }}(t^{\prime
})\}\rangle _{\rm{disc}}-2\langle \widehat{I}_{\eta }(t)\rangle \langle \widehat{I%
}_{\eta ^{\prime }}(t^{\prime })\rangle =0.
\end{equation}

\section{Recovering the Standard Formula for the Noise}

To prove Eq. (\ref{SLL_buttiker}) we note that the Green functions appearing in Eq. (\ref{SLL}) can be written as follow,
\begin{equation}
\mathbf{\G}^L \mathbf{G}^< = i \mathbf{\G}^L \mathbf{G}^r (n_L \mathbf{\G}^L + n_R \mathbf{\G}^R) \mathbf{G}^a,\label{d1}
\end{equation}
\begin{equation}
\mathbf{\G}^L \mathbf{G}^> = -i \mathbf{\G}^L \mathbf{G}^r [(1-n_L) \mathbf{\G}^L + (1-n_R) \mathbf{\G}^R] \mathbf{G}^a,\label{d2}
\end{equation}
\begin{equation}
\mathbf{\G}^L (\mathbf{G}^r-\mathbf{G}^a) = -i \mathbf{\G}^L \mathbf{G}^r (\mathbf{\G}^L + \mathbf{\G}^R) \mathbf{G}^a,\label{d3}
\end{equation}
\begin{eqnarray}
\mathrm{Tr}[\mathbf{\G}^L \mathbf{G}^a \mathbf{\G}^L \mathbf{G}^a &+& \mathbf{\G}^L \mathbf{G}^r \mathbf{\G}^L \mathbf{G}^r] = \mathrm{Tr} [2 \mathbf{\G}^L \mathbf{G}^r \mathbf{\G}^L \mathbf{G}^a +\phantom{x}\nonumber \\ && \phantom{xx} \mathbf{\G}^L (\mathbf{G}^r-\mathbf{G}^a) \mathbf{\G}^L (\mathbf{G}^r-\mathbf{G}^a)\label{d4}].
\end{eqnarray}
Now, defining the generalized transmission coefficients
\begin{eqnarray}
\mathbf{T}_{LL}&=&\mathbf{\G}^L \mathbf{G}^r \mathbf{\G}^L \mathbf{G}^a\\
\mathbf{T}_{LR}&=&\mathbf{\G}^L \mathbf{G}^r \mathbf{\G}^R \mathbf{G}^a,
\end{eqnarray}
we can write the above set of equations [Eqs. (\ref{d1})-(\ref{d4})] in terms of $\mathbf{T}_{LL}$ and $\mathbf{T}_{LR}$,
\begin{equation}
\mathbf{\G}^L \mathbf{G}^< = i n_L \mathbf{T}_{LL} +  i n_R \mathbf{T}_{LR},\label{d7}
\end{equation}
\begin{equation}
\mathbf{\G}^L \mathbf{G}^> = -i (1-n_L) \mathbf{T}_{LL} -i (1-n_R) \mathbf{T}_{LR},\label{d8}
\end{equation}
\begin{equation}
\mathbf{\G}^L (\mathbf{G}^r-\mathbf{G}^r) = -i \mathbf{T}_{LL}  -i \mathbf{T}_{LR},\label{d9}
\end{equation}
\begin{eqnarray}
\mathrm{Tr} [\mathbf{\G}^L \mathbf{G}^a \mathbf{\G}^L \mathbf{G}^a &+& \mathbf{\G}^L \mathbf{G}^r \mathbf{\G}^L \mathbf{G}^r] = \mathrm{Tr} [2 \mathbf{T}_{LL}-\phantom{xxxx}\nonumber \\ && (\mathbf{T}_{LL}+\mathbf{T}_{LR}) (\mathbf{T}_{LL}+\mathbf{T}_{LR})].\label{d10}
\end{eqnarray}
Using Eqs. (\ref{d7})-(\ref{d10}) in Eq. (\ref{SLL}) we obtain
\begin{eqnarray}
&& S_{LL}(\omega=0)=\frac{e^2}{\pi} \int d\e \mathrm{Tr} \{\nonumber \\ &&
n_L (1-n_L) \mathbf{T}_{LL} + n_L (1-n_R) \mathbf{T}_{LR} \nonumber \\ &&
+(1-n_L) n_L \mathbf{T}_{LL} +  (1-n_L)n_R \mathbf{T}_{LR} \nonumber \\ &&
+(n_L \mathbf{T}_{LL}  + n_R \mathbf{T}_{LR})[(1-n_L) \mathbf{T}_{LL} + (1-n_R) \mathbf{T}_{LR}] \nonumber \\  &&
-n_L (\mathbf{T}_{LL}  + \mathbf{T}_{LR})[(1-n_L) \mathbf{T}_{LL} + (1-n_R) \mathbf{T}_{LR}] \nonumber \\ &&
- (1-n_L) (\mathbf{T}_{LL}  + \mathbf{T}_{LR}) [n_L \mathbf{T}_{LL} + n_R \mathbf{T}_{LR}] \nonumber \\ &&
-n_L(1-n_L)[2 \mathbf{T}_{LL}- (\mathbf{T}_{LL}+\mathbf{T}_{LR}) (\mathbf{T}_{LL}+\mathbf{T}_{LR})] \}.\nonumber
\end{eqnarray}
The terms with $\mathbf{T}_{LL}$ cancel out identically, and the above expression can be written as
\begin{eqnarray}
&&S_{LL}(\omega=0)=\frac{e^2}{\pi}\int d\e
\nonumber\\&&\phantom{xxxx} \times \mathrm{Tr} \{
[n_L(1-n_L)+n_R(1-n_R)] \mathbf{T}_{LR}(\e)
\nonumber\\&&\phantom{xxxxxx}+ (n_L-n_R)^2 \mathbf{T}_{LR}(\e)
[1-\mathbf{T}_{LR}(\e)] \}.
\end{eqnarray}
Denoting $\mathbf{T}_{LR}$ simply as $\mathbf{T}$ we arrive at Eq. (\ref{SLL_buttiker}).


\begin{thebibliography}{99}
\bibitem{commentNina} This effect has been experimentally observed in Ref. [\onlinecite{cam07}].
\bibitem{diagram} A diagrammatic formulation for the shot noise in
a quantum dot coupled to ferromagnetic leads can be seen in [\onlinecite{mb06}].
\bibitem{comparison} Comparing the current plateaus obtained from Eq.
(\ref{ILwrearranged}) and the ones derived in Ref. [\onlinecite{at03}] we find a complete agreement between the results.
\end{thebibliography}
\end{document}